**Title**

**RECAP Framework v1.0: A Multi-Layer Meta-Architecture for Structuring Reasoning in Evidence Synthesis**

**Running Title**

**RECAP v1.0: Multi-Layer Reasoning Architecture**

**Author**

Hung Kuan Lee, MHS (candidate), MD

Department of Epidemiology

Johns Hopkins Bloomberg School of Public Health

Baltimore, MD, USA

**Corresponding Author**

Hung Kuan Lee

Email: **lhungku1@jh.edu**

Phone (Taiwan): **+886-975-250-176**

Address: Johns Hopkins Bloomberg School of Public Health, 615 N. Wolfe Street, Baltimore, MD 21205, USA

**Word Count**

(To be updated after full assembly) — *typically 7,000–12,000 for a foundational framework*

**Funding**

None.

**Conflicts of Interest**

The author declares no conflicts of interest.

**Data Availability**

Not applicable. No empirical data were used.

**Ethics Approval**

Not applicable.

# ABSTRACT

**Background:** Evidence synthesis has advanced substantially through improvements in reporting standards, bias assessment tools, and analytic methods, yet most workflows continue to rely on a single-layer structure in which conceptual, methodological, and procedural decisions coexist without clear boundaries [1–3,8,12]. As research programs expand across multiple projects, this structure produces unintentional divergence in definitions, measurement choices, and inferential pathways, contributing to cumulative inconsistency and methodological drift that are difficult to detect or reconcile [15,16]. Existing automation tools accelerate processing but do not provide a governing logic for maintaining coherence across projects [17,18].

**Objective:** RECAP Framework v1.0 aims to introduce a formal multi-layer meta-architecture that distinguishes stable methodological commitments from domain abstractions and project-level adaptations. The goal is to enable reasoning structures that are transparent, inheritable, and resistant to cross-project contamination.

**Methods:** The framework establishes a three-layer hierarchy—Grandparent, Parent, and Child—governed by strict inheritance and non-contamination rules. The Grandparent layer encodes methodological laws and philosophical constraints; the Parent layer translates these laws into domain-level structures without altering them; and Child modules instantiate individual projects while remaining insulated from influencing higher layers. Only abstract methodological insight may propagate upward, whereas domain content, measurement systems, and project-specific assumptions are structurally prohibited from doing so.

**Implications:** By formalizing abstraction boundaries and inheritance pathways, RECAP seeks to reduce unrecognized drift, improve internal coherence, and support cumulative methodological development across research programs. Version 1.0 represents an initial formulation intended to serve as a stable foundation for future refinements, domain-specific Parent engines, and automated Child-generation systems.

## 1. Introduction

### 1.1 Motivation

Contemporary evidence synthesis benefits from advances in reporting standards, structured appraisal tools, and analytic techniques, yet most workflows continue to treat methodological structure as something rebuilt from the ground up for every new project [8,12]. Teams repeatedly revisit definitions, measurement strategies, and inferential assumptions without a system capable of distinguishing which elements should remain stable across projects and which should appropriately vary. This pattern is not a failure of individual studies; it is a structural consequence of the single-layer model in which conceptual, methodological, and procedural decisions occupy the same undifferentiated space [1–3].

As research programs expand, gradual divergence in assumptions and reasoning pathways becomes inevitable. These divergences accumulate silently, producing inconsistencies that are difficult to detect from within any given project and nearly impossible to reconcile retrospectively [15,16]. Meanwhile, recent advances in automation accelerate processes such as screening, extraction, and risk-of-bias evaluation, but they do not supply the architecture needed to stabilize the inferential logic that guides these operations. In the absence of such architecture, automated systems risk reproducing the inconsistencies embedded in their inputs rather than mitigating them [17,18].

These structural limitations motivate the development of a framework that explicitly separates stable methodological commitments from project-level adaptations. RECAP Framework v1.0 is proposed as a multi-layer meta-architecture that achieves this separation. Its goal is not to replace existing tools or analytic techniques but to provide an organizing system in which methodological laws, domain abstractions, and project-level decisions can be distinguished, inherited, and evaluated with clarity [1,4]. By introducing these boundaries, RECAP seeks to reduce unrecognized drift, enhance internal coherence, and enable cumulative methodological development within and across research programs [15].

**1.2 Structural Problems Addressed by RECAP**

Most evidence workflows lack an abstraction layer that separates foundational methodological principles from domain-level conventions. Without such separation, conceptual and contextual decisions merge, obscuring which elements should remain fixed and which appropriately vary. This blending produces conceptual drift that typically remains invisible from within any single project [1–3,15].

A second structural limitation is the absence of a formal mechanism for methodological inheritance. In most research programs, each project reconstructs its assumptions independently; methodological insights rarely propagate reliably across studies [3,12]. Without a system for controlled inheritance, empirical conclusions may accumulate, but methodological understanding does not. A related problem is cross-project contamination, whereby assumptions developed within one context implicitly shape decisions in others without explicit justification. Such transfers can distort construct meanings and generate apparent inconsistencies across a body of work [15,16].

The increasing role of automation further magnifies these issues. Automated tools can dramatically accelerate evidence processing, but unless the underlying inferential structure is explicit, automation may scale inconsistencies rather than resolve them [17,18]. Together, these limitations demonstrate the need for a multi-layer framework capable of defining how methodological decisions are organized, stabilized, and inherited across projects.

**1.3 Positioning RECAP Within the History of Methodology**

The evolution of evidence synthesis has proceeded through recognizable phases: descriptive tabulation, structured review formats, standardized bias-assessment frameworks, and, more recently, integration of formal causal reasoning into review methodology [8,12]. These developments strengthened the rigor of individual analyses but retained the assumption that methodological decisions occur within a single project-level layer.

This long-standing assumption has produced an asymmetry in methodological progress: analytic tools and techniques have advanced, but the structural organization of reasoning across projects has not. No widely adopted framework currently governs how methodological principles relate to domain conventions or how project-level decisions inherit, refine, or diverge from broader structures.

RECAP Framework v1.0 addresses this gap by introducing a multi-layer meta-engine that governs the relationships among methodological laws, domain abstractions, and project implementations. To our knowledge, no existing methodological tradition articulates an inheritance-based architecture for multi-project evidence systems. RECAP should therefore be understood not as an incremental refinement of existing practice but as a structural extension to the methodological foundations of evidence synthesis.

**1.4 Rationale for a Multi-Layer Architecture**

Single-layer workflows require that conceptual definitions, methodological commitments, procedural rules, and analytic choices all be established within the same inferential space. Under this constraint, principles intended to remain stable across projects become entangled with decisions that should vary by context, making it difficult to determine which differences reflect legitimate adaptation and which represent conceptual drift [1–3].

A multi-layer architecture addresses this limitation by separating these decisions into distinct epistemic roles. Methodological laws reside in a domain-agnostic layer; domain abstractions translate these laws into families of constructs; and project modules instantiate specific studies within the constraints defined above. This stratification enables stability and flexibility to coexist in a disciplined manner.

The architecture also enables methodological inheritance. Insights generated within individual projects can be elevated in abstracted form to refine methodological principles, but contextual assumptions, measurement structures, and domain content cannot. This asymmetry ensures that cumulative refinement occurs without uncontrolled generalization of context-specific details [3,15]. At the same time, the architecture prevents cross-project contamination by restricting the movement of assumptions and definitions across projects unless explicitly elevated through appropriate channels.

**1.5 Contributions of the Framework**

RECAP Framework v1.0 contributes to evidence synthesis in five principal ways.

First, it establishes a multi-layer meta-architecture that separates methodological principles, domain abstractions, and project-level decisions. This separation clarifies the function of each decision type and prevents context-specific design choices from being mistaken for methodological laws.

Second, it formalizes methodological inheritance. Only abstractable insight may move upward; domain content, measurement structures, and project-specific assumptions may not. This mechanism supports cumulative methodological refinement while preventing uncontrolled drift [3,15].

Third, RECAP introduces explicit rules to prevent cross-project contamination. These rules ensure that assumptions originating within a particular context do not implicitly shape others unless reformulated at the appropriate level of abstraction.

Fourth, it provides a structure capable of integrating automation without sacrificing conceptual coherence. Automation can accelerate many components of evidence synthesis, but without structural boundaries it may amplify inconsistencies [17,18]. RECAP offers a principled architecture within which automation can operate safely.

Finally, RECAP aims to establish a foundation for the long-term evolution of evidence systems. By supplying rules for inheritance, separation, and methodological refinement, it supports coherence across generations of research rather than within isolated projects alone.

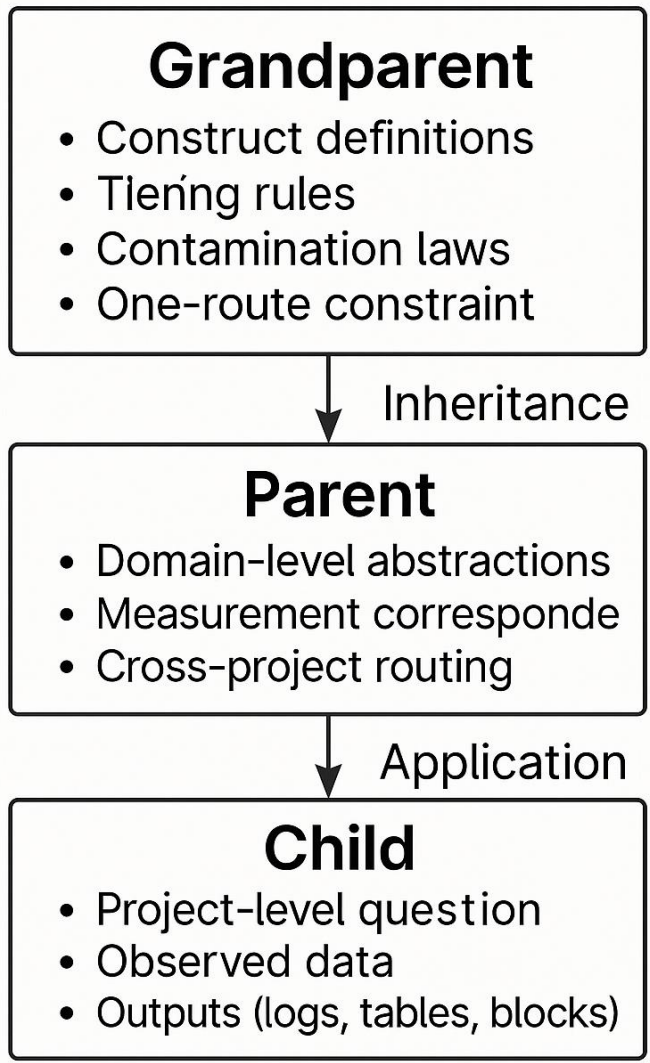

RECAP three-layer architecture

## 2. Theoretical Foundations

### 2.1 Scientific Philosophy Underlying RECAP

RECAP Framework v1.0 rests on three philosophical commitments that define its epistemic boundaries and shape its structural design.

The first is an **anti-reification principle**, grounded in long-standing critiques within philosophy of science and measurement theory [1–4]. Constructs, categories, and variables are treated as analytic instruments rather than natural entities. Their legitimacy arises from their inferential function—not from any inherent status. This stance prevents empirical artifacts from being mistaken for theoretical objects and protects the architecture from interpretive drift as projects accumulate.

The second commitment is **multi-level reasoning**. Scientific inference occurs across different strata of abstraction, each subject to distinct justificatory requirements [1–3]. When all decisions are made within a single undifferentiated space, distinctions between methodological law, domain-level convention, and project-level implementation are blurred. By explicitly separating these layers, RECAP ensures that each decision is evaluated within the epistemic register appropriate to its role.

The third commitment is **model-bound meaning**, which holds that constructs derive their meaning from their location within the inference architecture, not from isolated definitions. A construct situated at the methodological layer functions differently from one situated in a domain abstraction or project-level implementation. This position dependence prevents constructs from migrating across contexts without explicit re-specification, thereby stabilizing interpretation across successive projects.

Together, these commitments provide the philosophical grounding for RECAP's inheritance system and its strict separation between layers.

**2.2 RECAP as a Theory: Core Postulates**

RECAP Framework v1.0 is not merely a set of recommendations; it is a theory of how multi-project evidence systems must be structured to avoid drift. Its theoretical identity is defined by seven core postulates.

**Postulate 1: Construct–Measurement Separation**
Constructs must be defined at the conceptual level and insulated from any particular operationalization. Measurements approximate constructs but never define them. This principle prevents constructs from collapsing into empirical proxies.

**Postulate 2: One Route per Project**
A project must commit to exactly one inferential route. Competing inferential logics cannot be blended without undermining interpretability. The one-route rule provides epistemic discipline and prevents hybrid reasoning.

**Postulate 3: Upward Transmission of Insight Only**
Only abstract methodological insight may move upward in the inheritance hierarchy. Domain content, measurement choices, and project-specific assumptions may not. This asymmetry ensures that methodological laws evolve without absorbing contextual detail.

**Postulate 4: The Meta-Engine Is Permanently Abstract**

Methodological laws at the Grandparent layer cannot be rewritten from below. Lower layers may refine the abstraction through insight, but they cannot modify its structure. This protects the universality and stability of the framework.

**Postulate 5: All Inferences Require a Disconfirming Model**

Every inferential claim must be accompanied by at least one plausible alternative that could challenge or overturn it. This postulate aligns RECAP with falsificationist principles [1–3] and prevents over-extension of claims.

**Postulate 6: Transparency Is Structural**

A project operating under RECAP must produce a Study Log, a Tier Table, and a Reviewer Block. These are not reporting preferences but structural requirements enabling reconstructability and reviewer-critical reasoning.

**Postulate 7: Inheritance Is Directional**

Constraints flow downward; methodological insight flows upward; lateral or reverse transfers of domain content are prohibited. Directionality preserves coherence across multi-project ecosystems.

These postulates jointly define RECAP as a theory of governance for evidence systems rather than an analytic method or procedural tool.

## 2.3 Falsifiability and Boundary Conditions

A methodological framework is meaningful only if it can be falsified. RECAP provides several opportunities for challenge.

First, if the multi-layer architecture were shown to induce systematic bias relative to simpler designs, its conceptual advantage would be undermined. Second, if a single-layer workflow could reliably maintain conceptual stability across projects without drift, the rationale for structural separation would weaken. Third, RECAP would be challenged if controlled contamination—normally prohibited within the framework—were shown to enhance coherence or interpretability rather than degrade it.

The framework also delineates clear boundary conditions.
A multi-layer architecture is not required for:

- small, self-contained projects,
- settings where constructs, measurements, and designs are tightly coupled,
- or environments where long-term inheritance is irrelevant.

RECAP's advantages emerge most clearly in large, long-running, or multi-domain research programs where uncontrolled interactions among projects introduce interpretive ambiguity or conceptual drift [15,16]. In such environments, the costs of structural governance are outweighed by gains in stability, transparency, and cumulative methodological learning.

**2.4 Relation to Existing Theories of Scientific Reasoning**

RECAP builds upon long-standing traditions in scientific philosophy while filling gaps not addressed by prior frameworks.

Its anti-reification stance echoes foundational work emphasizing the constructed nature of scientific variables and the need for conceptual discipline [1–4]. Its requirement for disconfirming models aligns with falsificationist reasoning, which holds that the strength of an inference depends on the strength of the alternatives it survives [1–3].

Where RECAP diverges is in its **structural orientation**.

Most accounts of scientific inference—whether based on causal modeling, hypothesis testing, or probabilistic reasoning—operate within a single analytic frame. They rarely address how reasoning should be governed **across projects**, or how methodological insights should accumulate **over time**.

Similarly, hierarchical models in the philosophy of explanation distinguish between levels of abstraction but do not specify rules for how layers should interact when evidence is distributed across diverse projects. RECAP operationalizes these distinctions through explicit inheritance rules, boundary conditions, and contamination controls.

In this sense, RECAP extends classical theories of scientific reasoning into the domain of evidence ecosystems: environments in which many projects co-exist, interact, and evolve across generations.

**2.5 Why Evidence Systems Require Governance Layers**

Evidence synthesis traditionally focuses on producing valid inferences within individual projects. For isolated analyses, well-established reporting standards and methodological conventions often suffice to maintain internal coherence [8,12]. However, as research programs expand, small divergences in definitions, assumptions, and analytic choices begin to accumulate. These divergences may each be defensible in isolation yet collectively produce conceptual drift that is invisible within any single project [15,16].

A governance layer addresses this problem by defining how decisions are separated,

constrained, and inherited. Without such governance, multi-project programs resemble loosely connected families of analyses whose conceptual relationships are informal rather than principled. This informality becomes a liability in environments where:

- multiple constructs evolve over time,
- analytic decisions propagate implicitly through teams,
- datasets differ in structure or measurement, or
- results from one project are used to justify another.

Governance layers are especially critical when automation is introduced. Automated tools can accelerate screening, extraction, and appraisal, but such tools do not understand conceptual boundaries. Automation amplifies whatever structure it is given; without a meta-architecture, it may spread inconsistencies rather than correct them [17,18].

By enforcing separation of layers, controlling information flow, and regulating how insights accumulate, RECAP supplies the structural stability that large-scale, multi-project evidence ecosystems require. It is not designed merely to improve individual reviews, but to govern the intellectual lineage through which evidence systems evolve.

## 3. Definitions

This section establishes the foundational vocabulary that governs all RECAP-based systems. These definitions are not descriptive labels but constitutional elements of the meta-architecture. They determine how constructs, evidence, and reasoning may operate within RECAP and how they are prevented from drifting across layers or projects.

Each definition is layer-anchored: a term's meaning is fixed by its position in the inheritance structure and cannot migrate without formal re-specification.

**Construct**

A **construct** is a purely conceptual target of inference. It exists only at the level of abstraction and is not reducible to any particular dataset, instrument, or operationalization. Constructs are defined independently of measurement in order to preserve interpretive stability across projects and iterations.

Within RECAP, constructs are *non-collapsible*: their meaning does not change when measurements change. The construct anchors the inferential objective, whereas measurements approximate but do not define it.

**Measurement**

A **measurement** is an operational approximation of a construct. It may be derived from instruments, coding procedures, classification rules, or analytic transformations. Measurements are regarded as contingent, fallible, and replaceable; their epistemic status derives entirely from how well they approximate the construct. Measurement does not possess independent authority. Changes in measurement do not constitute changes in the construct.

**Tier Core**

**Tier Core** contains evidence units whose construct alignment, measurement adequacy, and design structure satisfy all requirements of the project's declared inferential route. Units in Core provide the primary evidential channel. Core designation does *not* imply high quality, causal validity, or statistical strength. It reflects only structural compatibility with the route's assumptions and conceptual definitions. Core determines *which* evidence is allowed to shape the primary inference.

**Tier Supplement**

**Tier Supplement** includes evidence units that are relevant to the inferential objective but constrained by limitations in measurement, design, or reporting.

Supplement units fulfill three epistemic roles:

1. **Boundary articulation** — clarifying the outer limits of the inference.

2. **Robustness evaluation** — testing whether conclusions remain stable under alternative conditions.

3. **Contextualization** — supplying interpretive scaffolding without determining the primary claim.

Supplement is not a discard category; it is a *structurally secondary* one.

**Tier Excluded**

**Tier Excluded** contains evidence units that cannot support meaningful inference under the declared route. Reasons include:

- fundamental construct mismatch,

- measurement failures that cannot be reconciled with the construct,

- design characteristics that violate non-negotiable assumptions, or

- irreducible opacity in reporting.

Exclusion does not remove a unit from documentation. It establishes the epistemic perimeter of the search space and preserves transparency.

**Route**

A **route** is a single, pre-declared inferential pathway specifying how evidence will be interpreted under a coherent set of assumptions. A project may commit to only *one* route. This prevents hybrid reasoning, post-hoc assumption mixing, and conceptual ambiguity.

A route defines:

- the construct to be inferred,
- the assumptions enabling inference,
- the allowable forms of evidence, and
- the disconfirming models that bound the claim.

Route commitment is a constitutional requirement of RECAP.

**Grandparent / Parent / Child Layers**

These layers define the architecture of inheritance governing all RECAP systems:

**Grandparent Layer**

Contains universal methodological laws.

These laws are permanently abstract, domain-agnostic, and cannot be rewritten from below. They function as the constitution of the meta-engine.

**Parent Layer**

Translates methodological laws into domain-level abstractions.

The Parent layer defines domain constructs, permissible measurement categories, and allowed design forms. Parents govern families of Child projects.

Children may not modify Parent definitions.

**Child Layer**

Implements project-level reasoning.

Each Child contains its own route, its own tiering, its own Study Log, and its own Reviewer Block. Children inherit constraints from Parents and Grandparents but cannot rewrite them.

This hierarchy ensures directional inheritance: stability above, flexibility below.

**Contamination (Upward, Downward, Horizontal)**

**Contamination** refers to unlawful movement of assumptions, content, or definitions across layers in violation of RECAP's inheritance rules.

- **Upward contamination**: contextual or domain-specific material attempts to rewrite methodological laws. (Most severe)

- **Downward contamination**: a project implicitly alters constraints that should govern it.

- **Horizontal contamination**: assumptions migrate informally between Child projects.

Contamination undermines coherence and nullifies RECAP's structural guarantees.

**Insight Transmission**

**Insight transmission** is the controlled process by which a methodological insight—never content—moves upward in abstraction.

To be eligible for transmission:

1. the insight must be independent of domain content,

2. it must be expressible in the conceptual vocabulary of the receiving layer, and

3. it must refine, not rewrite, the methodological abstraction.

Transmission allows RECAP to evolve across generations without inheriting contextual noise.

**Evidential Unit (Unit of Tiering)**

An **evidential unit** is the smallest analysable entity that can be independently evaluated for tiering. It may represent:

- a whole study,

- a study arm,

- a construct-specific component, or

- a subgroup observable.

The unit must be pre-specified in the Study Log. Tiering always applies to units, not to "studies" in the vague sense.

**Assumption**

An **assumption** is a logical or methodological precondition required for an inferential route to hold. Assumptions must be explicitly declared, justified, and bounded by at least one disconfirming model. Undeclared assumptions constitute structural violations under RECAP.

**Reviewer Block**

A **Reviewer Block** is a structured reflection tool that reproduces the scrutiny of a rigorous reviewer. It documents:

- at least two methodological findings,
- one conceptual insight,
- one anticipated critique,
- one disconfirming model,
- the assumptions required for the declared route.

The Reviewer Block enforces internal accountability before external review.

**Study Log**

The **Study Log** is the canonical archive of the evidence universe for a project. It contains tier assignments, justifications, measurement notes, bias considerations, and record of any contamination or reclassification events. Without a Study Log, a project cannot be reconstructed and therefore cannot be considered RECAP-compliant.

**Tier Table**

The **Tier Table** summarizes only Core and Supplement units and provides a structured account of their strengths, limitations, and methodological relevance. It links the inferential logic of the route to the evidence that supports it.

**4. Three-Layer Reasoning Architecture**

The RECAP Framework organizes all inferential activity into a three-layer architecture. These layers—**Philosophical Rigor**, **Reviewer-Critical Reasoning**, and **Adaptive Representation**—are not sequential steps but co-governing strata, each with a non-interchangeable epistemic function. Their separation prevents reasoning drift, maintains methodological stability across projects, and ensures that communication pressures never reshape inferential structure.

The three layers are governed by the inheritance hierarchy established in Section 3:
Layer A corresponds to Grandparent (methodological laws),
Layer B to Parent-level enforcement and scrutiny,
Layer C to Child-level communication and interface.

This mapping ensures that conceptual, methodological, and representational decisions occupy distinct epistemic spaces.

**4.1 Layer A — Philosophical Rigor (Foundational Constraints)**

Layer A defines the non-negotiable constraints that govern all reasoning within a RECAP system.
These constraints do not depend on domain, dataset, or analytic design. Their function is to maintain conceptual integrity across generations of projects.

Layer A enforces three foundational commitments:

### 4.1.1 Construct–Measurement Separation

Constructs are conceptual; measurements approximate them.

Layer A prohibits any collapse of construct meaning into its operationalization. This separation ensures stability across studies, replications, and domains, and prevents methodological drift when measurement tools evolve.

### 4.1.2 Anti-Reification

Variables, categories, and indicators are treated as analytic instruments rather than natural kinds. Their legitimacy derives from their inferential role, not their labels or their historical use.

This prevents operational conveniences from being mistaken for ontological truths.

### 4.1.3 Disconfirming Models as a Structural Obligation

Every inferential claim must be bounded by at least one plausible alternative model capable of undermining or reversing the conclusion.

This requirement operationalizes falsifiability and ensures that every inference is proportionate to its evidential environment.

**Purpose of Layer A:**

To provide a philosophical constitution for reasoning—rigorous enough to remain invariant, abstract enough to govern any domain, and constrained enough to prevent conceptual drift even when automation scales evidence processing.

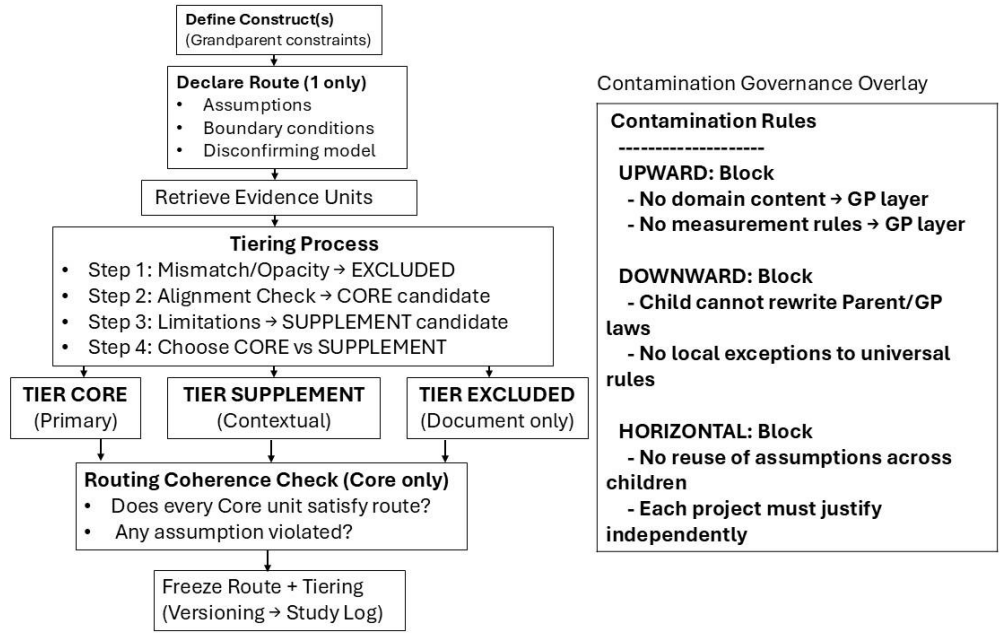

## 4.2 Layer B — Reviewer-Critical Reasoning (Structured Methodological Scrutiny)

Layer B operationalizes the posture of a rigorous methodological reviewer.

Where Layer A establishes universal constraints, Layer B enforces disciplined reasoning within the boundaries of a declared route.

Layer B requires explicit scrutiny of the following:

### 4.2.1 Coherence Between Constructs and Measurements

Measurements must approximate constructs without distorting their conceptual meaning. Misalignment must be documented and addressed in the Study Log or Reviewer Block.

### 4.2.2 Bias and Its Directional Implications

Bias sources must be described not only in presence/absence terms but with explicit directional hypotheses—how each bias could alter magnitude, sign, or interpretability.

### 4.2.3 Alternative Explanations

Layer B requires articulation of all plausible rival explanations consistent with the evidence, regardless of whether they are quantified.

### 4.2.4 Constraint Matching Between Evidence and Route

Every Tier Core unit must satisfy the assumptions of the declared route without speculative reconstruction.

Supplement units may satisfy assumptions conditionally; Excluded units may not satisfy them at all.

### 4.2.5 Proportionality Between Claim and Support

Claims must not exceed the inferential authority provided by Core evidence and must be explicitly framed relative to disconfirming models.

**Purpose of Layer B:**

To ensure reconstructable, internally coherent reasoning that would withstand adversarial peer review.

Layer B is not a reporting guideline or checklist—it is an epistemic enforcement mechanism.

## 4.3 Layer C — Adaptive Representation (Communication Without Drift)

Layer C governs how reasoning is expressed to external audiences, including journals, reviewers, policy stakeholders, and interdisciplinary readers.

Crucially, Layer C is representational, *not inferential*. It may translate, summarize, restructure, or re-emphasize—but may never reinterpret.

### 4.3.1 What Layer C May Modify

- Lexical density
- Narrative flow
- Placement of constructs, results, or conceptual framing
- Integration of figures, tables, or graphical supplements
- Style requested by a journal or discipline

These modifications affect communication efficiency but not the reasoning architecture.

### 4.3.2 What Layer C May Not Modify

Layer C may not:

- alter construct definitions,
- relax or revise route assumptions,
- promote Supplement evidence into Core,
- introduce claims not supported in Layer B,
- weaken or strengthen the evidence beyond what the route allows.

Layer C is structurally insulated from influencing inferential logic.

**Purpose of Layer C:**

To allow projects to adapt to diverse communication contexts while preserving the epistemic commitments of Layers A and B.

### 4.4 Why the Three-Layer Structure Is Necessary

Traditional single-layer workflows collapse philosophical constraints, methodological reasoning, and journal-driven communication into a single plane. This produces predictable structural failures:

1. **Reporting styles reshape reasoning**, as authors modify claims to satisfy reviewer preferences.

2. **Project-level assumptions masquerade as general methodological principles**, generating long-term drift.

3. **Measurement convenience becomes mistaken for construct definition**, especially in automated settings.

4. **Heterogeneity across projects accumulates silently**, obscuring interpretability of a research program.

By assigning each class of decision to its own epistemic layer, RECAP:

- protects methodological laws from contextual distortion,

- allows project-level flexibility without rewriting the meta-engine,

- ensures automation cannot alter conceptual boundaries,

- and maintains coherence across multi-project, multi-domain, or long-running evidence ecosystems.

**4.5 How the Layers Co-Govern Projects**

The layers operate jointly:

- **Layer A** defines the permissible universe of inference.

- **Layer B** determines the validity of reasoning within that universe.

- **Layer C** governs how the reasoning is externally communicated.

No layer can substitute for another.

Their functional separation is the mechanism by which RECAP preserves methodological stability even when projects proliferate or teams change.

### 4.6 Boundary Conditions

RECAP's three-layer architecture is essential when:

- research programs involve multiple projects over time,

- constructs evolve across contexts,

- automation plays a significant role in screening or synthesis, or

- teams require stable methodological inheritance.

For small, single-use analyses with narrow scope, a reduced form of layering may be sufficient, but the decision to simplify must itself be documented as a methodological choice.

### 5. Tiering System

The tiering system is RECAP's structural mechanism for organizing evidence according to its compatibility with the declared inferential route.

Unlike risk-of-bias tools or grading schemes, tiering does **not** evaluate effect direction, magnitude, or internal validity.

Its purpose is to determine which evidence can legitimately participate in the inferential

structure — and under what epistemic role.

Tiering is therefore a **governance function**, not a quality filter.

It defines the evidential backbone of a project and prevents conceptual drift as research programs accumulate across time.

**5.1 Purpose and Function of Tiering**

Tiering ensures four foundational requirements within any RECAP-based project:

1. **Construct coherence**

   Evidence must align with the conceptual definition declared by the route.

2. **Measurement adequacy**

   Operational indicators must approximate the construct without distorting its meaning.

3. **Design compatibility**

   The study's structure must satisfy the assumptions required by the route.

4. **Boundary control**

   Evidence that cannot satisfy route assumptions is prevented from silently influencing the main inference.

Tiering is the system that maintains a disciplined evidential universe and protects the inferential stream from contamination.

**5.2 Tier Core**

Tier Core contains evidence units capable of supporting **primary inference** under the designated route.

A unit qualifies for Tier Core only when:

1. **Construct alignment**

   Its conceptual target matches the project's declared construct without reinterpretation.

2. **Measurement plausibility**

   Its operationalization provides a credible approximation of the construct.

3. **Design sufficiency**

   Its design and reporting allow inference under the route's assumptions without speculative reconstruction.

Tier Core forms the **inferential surface** of the project — the set of units from which claims may be drawn without qualification.

Core membership is **structural privilege**, not a judgment of superiority.

**5.3 Tier Supplement**

Tier Supplement contains units that carry **relevant but methodologically constrained** information.

A unit is routed to Supplement when:

- construct alignment is adequate,

- measurement approximates the construct imperfectly,

- or design/reporting limitations prevent participation in primary inference.

Supplement units serve three epistemic roles:

1. **Boundary assessment**

    They reveal the limits of the primary inference.

2. **Robustness evaluation**

    They test whether conclusions remain stable under alternative assumptions.

3. **Contextual interpretation**

    They provide interpretive depth without determining the claim.

Supplement evidence is **structurally secondary**, not disposable.

**5.4 Tier Excluded**

Tier Excluded contains evidence units that cannot participate in inference because of:

- fundamental construct mismatch,

- measurement that fails to approximate the construct,

- design features incompatible with non-negotiable route assumptions,

- or insufficient information that would require speculation to interpret.

Exclusion is not erasure.

Every excluded unit remains documented in the Study Log to preserve transparency and to delineate the epistemic boundary of the evidence base.

Excluded classification is **structural, not normative**.

**5.5 Tiering Algorithm (RECAP v1.0)**

Tiering follows a fixed, non-circumventable sequence:

**Step 1 — Detect fundamental mismatch or opacity**

If the construct cannot be reconciled with the declared definition or essential information is absent, assign **Tier Excluded** immediately.

**Step 2 — Identify candidates for Core**

Evaluate construct alignment, measurement plausibility, and design compatibility.

**Step 3 — Assess methodological limitations**

If conceptually aligned but limited (measurement mismatch, partial reporting, design constraints), assign **Tier Supplement**, unless limitations make inference impossible.

**Step 4 — Resolve Core vs Supplement**

Core is reserved for units requiring no substantial qualification.

Supplement is for units requiring conditional assumptions.

**Step 5 — Document justification**

Tiering without justification is non-compliant.

Each decision must reference construct alignment, measurement adequacy, and design considerations.

**5.6 Handling Ambiguity**

RECAP imposes three rules:

**Rule 1 — No speculative reconstruction**

If inference requires assumptions not anchored in the reported data → **Tier Excluded**.

**Rule 2 — Ambiguity resolvable through explicit assumptions → Tier Supplement**

Supplement absorbs units requiring conditional interpretation.

**Rule 3 — Multiple interpretations**

Split the unit when possible.

If unsplittable, assign to the **more conservative tier**.

These rules prevent hidden assumptions from contaminating the inferential stream.

**5.7 Re-Tiering and Versioning**

Re-tiering is allowed only when new information becomes available.

Each reclassification must include:

- timestamp,

- source of new information,

- justification for tier revision,

- implications for the inferential route.

Silent modification is prohibited because it destroys genealogical traceability.

Reviewer disagreements must be logged and summarized in the Reviewer Block.

**5.8 What Tiering Does *Not* Decide**

Tiering does **not**:

- assess risk of bias,

- determine effect direction or magnitude,

- establish causal interpretation,

- replace appraisal tools,

- imply evidential "quality."

Tiering governs **structural alignment**, not statistical or causal validity.

**5.9 Boundary Conditions for Tiering**

Tiering is required when:

- a program contains multiple related projects,

- constructs evolve across contexts,

- automation interacts with evidence streams,

- long-term coherence is necessary.

A simplified tiering system may be used for narrow, one-off analyses only if explicitly justified. Regardless of scope, **exclusion always requires documentation**. Granularity must reflect meaningful distinctions: over-fragmentation weakens interpretability; under-fragmentation collapses essential structure.

**6. Routing System**

Routing is RECAP's mechanism for governing *how inference is constructed*. Where tiering organizes **what evidence may be used**, routing organizes **how the evidence is allowed to mean anything at all**.

In single-layer workflows, this inferential logic is often implicit: constructs, assumptions, and analytic operations co-exist in an undifferentiated decision space.

RECAP rejects this structure.

A project cannot claim coherence if it cannot articulate the pathway linking its constructs to its conclusions.

Routing formalizes this pathway, transforming informal intuition into a **governed inferential architecture**.

Routing has two irreducible properties:

1. **A project must commit to exactly one route.**

2. **A route must fully determine how Tier Core evidence will be interpreted.**

Without routing, tiering has no epistemic target, and inference collapses into ad hoc justification.

**6.1 The One-Route Principle**

The one-route requirement is not a stylistic restriction — it is a **structural necessity**.

Each inferential route embodies a distinct constellation of:

- construct definitions,

- admissible assumptions,

- boundary conditions,

- plausible disconfirming models,

- and interpretive constraints.

If a project were allowed to operate under multiple routes simultaneously, it would inhabit mutually incompatible epistemic spaces. Assumptions could be blended post hoc, enabling selective interpretation and undermining falsifiability.

Thus:

**A project may compare routes, BUT may never operate under more than one.**

Comparison is exploratory.

Commitment is inferential.

The one-route principle is the mechanism that prevents:

- hidden assumption mixing,

- switching logics after seeing results,

- conceptual drift,

- and interpretive ambiguity across large programs.

It is the rule that turns inference into a *governed system* rather than a sequence of decisions.

**6.2 Routing Algorithm (v1.0)**

A route is not a preference — it is a **procedural declaration** that binds the project.

Below is the canonical routing sequence under RECAP:

**ROUTING ALGORITHM (v1.0)**

1. **DEFINE the target construct(s).**
   Clarify the conceptual object of inference without reference to measurement.

2. **DECLARE the inferential objective.**

   Examples: comparative, prognostic, descriptive, stability-mapping.

3. **SPECIFY the assumptions necessary for this objective.**

   e.g., exchangeability, temporal ordering, measurement validity, independence constraints.

4. **ENUMERATE all permissible inferential routes** consistent with Steps 1–3.

5. **SELECT exactly one route** based on methodological fit — not convenience.

6. **DOCUMENT the chosen route** in the Study Log, including:

   - formal construct definition,
   - required assumptions,
   - anticipated failure modes,
   - disconfirming models.

7. **RETRIEVE evidence units and perform tiering.**

   Tiering cannot occur without prior route declaration.

8. **FILTER evidence into the inferential stream:**

   - **Core** units → primary inference
   - **Supplement** → sensitivity / boundary / contextual roles
   - **Excluded** → documented but non-inferential

9. **CHECK coherence** between route assumptions and Core evidence.

If Core evidence cannot sustain the route → the route must be revised (with versioning).

10. **FREEZE the route** prior to analysis.

    Any revision requires justification, timestamping, and documentation of downstream implications.

Interpretation:

Steps 1–4 define the conceptual universe; Step 5 forces commitment; Step 6 preserves transparency; Steps 7–10 prevent retrospective reinterpretation.

Routing, once frozen, becomes the **epistemic contract** of the project.

**6.3 Transparency Requirements**

Routing is only meaningful when transparently documented.

A RECAP-compliant project must make the following components public:

1. **Construct definitions** and their non-collapsible conceptual meaning.

2. **Assumptions enabling inference**, including where they may break.

3. **Alternative routes considered but rejected**, with rationale.

4. **Mapping between tier assignments and the chosen route.**

5. **Versioned record of all changes**, including corrections or re-tiering triggered by new information.

Without transparency, routing collapses into ad hoc decision-making and loses its methodological integrity.

### .4 What Routing *Does Not* Do

Routing does **not**:

- dictate statistical models,
- determine effect sizes,
- evaluate bias or confounding,
- impose reporting structure,
- or rank evidence.

Routing is **not an analytic tool** — it is the governance mechanism that determines what *counts* as a valid inference.

It defines the permissible inferential universe.

Modeling, estimation, and interpretation occur **within** that universe.

### 6.5 Boundary Conditions for Routing

Routing becomes essential in:

- multi-project evidence ecosystems,
- automated workflows that require pre-specified architecture,
- programs where constructs evolve across time,
- settings requiring reproducible interpretive boundaries.

Routing may be simplified only in narrowly scoped, single-study analyses — and even then, the one-route declaration must still be made.

When omitted, reasoning reverts to the single-layer model, where assumptions drift silently across decisions and across generations of research.

**6.6 Why Routing Improves Inferential Stability**

Routing prevents three systemic failure modes:

1. **Assumption drift**

   Without a route, assumptions migrate unconstrained across steps.

2. **Hybrid logic**

   Projects inadvertently mix incompatible inferential frameworks.

3. **Retrospective reinterpretation**

   Decisions are reconstructed after results are known, weakening falsifiability.

By enforcing a single, frozen inferential pathway, routing generates:

- reproducibility,
- coherence,
- explicit limits on inference,
- and long-term interpretability across a research program.

Routing is the mechanism that converts a project from a set of analytic steps into a **disciplined inferential structure**.

**7. Contamination Control System**

Layer separation is the defining structural feature of the RECAP Framework. Without strict control of how information flows across layers, the distinctions between

**methodological law, domain abstraction, and project-level implementation** collapse.

When this collapse occurs, RECAP does not merely degrade — it ceases to exist.

Contamination is therefore not a minor error but a **category mistake** that threatens the epistemic identity of the system.

RECAP formalizes three classes of contamination:

1. **Upward contamination**（most severe）
2. **Downward contamination**
3. **Horizontal contamination**

Each represents a distinct and diagnosable failure mode in multi-project evidence ecosystems.

**7.1 Upward Contamination (Most Severe)**

**Definition:**
Upward contamination occurs when **content, assumptions, or measurement conventions** originating in a Parent or Child layer attempt to modify principles in a higher layer — especially the Grandparent layer.

**Why it is severe:**
Upward contamination forces the meta-engine to absorb empirical or contextual information, collapsing abstraction into specificity.
This destroys its ability to govern multiple domains and breaks the inheritance structure.

**Examples:**

- Using a domain-specific construct (e.g., "rapid cycling BD") to redefine a Grandparent-level concept (e.g., "construct").

- Allowing a measurement instrument (e.g., PHQ-9, OCT) to reshape methodological laws.

- Importing assumptions from a specific dataset into the Parent layer's domain abstractions.

**Outcome:**

The entire architecture becomes non-generalizable — it becomes a *case study*, not a framework.

Upward contamination is the most severe violation because it rewrites the constitution of the system.

**7.2 Downward Contamination**

**Definition:**

Downward contamination occurs when a Child module implicitly rewrites or weakens methodological or domain-level laws to accommodate local convenience or analytic preference.

**Why it matters:**

Downward contamination fractures the coherence of the research ecosystem by allowing project-specific exceptions to override universal constraints.

**Examples:**

- Relaxing the one-route principle because a dataset appears to support multiple

interpretations.

- Allowing construct–measurement collapse ("we will treat this proxy as the construct") to justify a preferred analytic model.

- Introducing domain-level exceptions ("in glaucoma, we don't apply that rule") that contradict the Parent layer.

**Outcome:**

Inference becomes locally coherent but globally incoherent — each project becomes its own methodological universe.

Downward contamination destroys cumulative methodological progress.

### 7.3 Horizontal Contamination

**Definition:**
Horizontal contamination refers to assumptions, measurement interpretations, or analytic heuristics migrating from one Child project to another without explicit routing or justification.

**Why it is dangerous:**
Horizontal contamination is subtle, common, and often goes unnoticed — but it produces cross-project drift that cannot be traced or corrected.

**Examples:**

- Replicating measurement coding rules from one project to another without re-evaluating construct alignment.

- Adopting a study design convention used in a different phenotype or domain

because "it worked before."

- Allowing terminology to carry inferential meaning across unrelated projects.

**Outcome:**

Projects begin to share assumptions informally, collapsing the independence required by the inheritance structure.

Horizontal contamination is often the first sign that a research program has outgrown a single-layer workflow.

**7.4 Allowed vs. Prohibited Information Flow**

To prevent contamination, RECAP functions as an **information-gated system**.

**Allowed Flows (Strictly Controlled)**

- **Grandparent → Parent → Child**

    (constraints flow downward)

- **Child → Parent**

    *(insight only, abstracted and re-expressed at higher level)*

**Prohibited Flows (Always Violations)**

- **Parent → Grandparent**

    (attempt to rewrite methodological law)

- **Child → Parent/Grandparent**

    (attempt to elevate content or assumptions)

- **Child ↔ Child**

without explicit justification (assumption transfer)

The framework remains stable only if these flows are enforced.

**7.5 Contamination Detection Algorithm (v1.0)**

RECAP provides a procedural system for identifying and correcting contamination events.

**CONTAMINATION DETECTION ALGORITHM（v1.0）**

1. **IDENTIFY source**

   Determine whether the information originated in Grandparent, Parent, or Child layers.

2. **CLASSIFY the type**

   - content
   - measurement
   - assumption
   - methodological insight

3. **CHECK flow permissions**

   Compare the actual direction of movement to the allowed flow map.

4. **FLAG the contamination**

   If it violates allowed flows, classify as Upward / Downward / Horizontal.

5. **TRACE downstream implications**

   Identify which tier assignments, routes, or claims may be affected.

6. **DOCUMENT in Study Log / Parent Log**

    Required fields:

    - rule violated
    - nature of contamination
    - risks introduced
    - decisions affected
    - corrective action

7. **QUARANTINE or REVERSE**

    Remove or isolate the contaminated component.

8. **EVALUATE for abstract insight**

    If the contamination reveals a structural issue (e.g., inadequate domain abstraction), extract methodological insight and re-express it at the Parent or Grandparent level.

9. **VERSION the correction**

    Update the framework with a timestamped entry describing the event.

The algorithm ensures contamination cannot silently influence inference.

## 7.6 Boundary Conditions

Full contamination control is **essential when**:

- multiple Child projects share a Parent layer,
- automation is used to generate or process evidence streams,

- constructs evolve over time,

- researchers rotate across projects,

- large teams collaborate on long-running programs.

Contamination control may be simplified only in isolated, one-off analyses.

Even then, the system requires explicit justification for bypassing full enforcement.

**7.7 Structural Guarantees Against Drift**

RECAP prevents drift through five guarantees:

1. **Layer separation**

    Each type of reasoning is confined to its epistemic layer.

2. **One-route constraint**

    Inference is never constructed from multiple incompatible pathways.

3. **Versioned tiering and routing**

    Decisions cannot mutate silently.

4. **Documentation of contamination events**

    The genealogy of reasoning remains visible.

5. **Restricted upward insight flow**

    Only abstract methodological insight may migrate upward.

Together, these guarantees allow RECAP to function as a **multi-generational methodological engine** rather than a one-cycle tool.

**8. Mandatory Study-Level Outputs**

The RECAP Framework requires every Child project to generate a minimum set of outputs that make its inferential structure *auditable, reconstructable,* and *inheritance-compatible*.

These outputs are not reporting conveniences nor templates—they are **structural obligations** that operationalize RECAP's transparency and governance principles.

Without these outputs, a project cannot claim compliance with RECAP, regardless of how carefully it tiers evidence or documents its route.

The mandatory outputs are:

1. **Study Log** (unit-level audit record)

2. **Tier Table** (alignment summary for Core + Supplement units)

3. **Reviewer Block + Analytic Memo** (Layer B internal scrutiny)

Together, these components function as the project's epistemic ledger—the permanent record of how decisions were made, justified, and structurally constrained.

**8.1 Study Log (Canonical Audit Record)**

The **Study Log** is the foundational accountability mechanism of RECAP.
It documents *every* evidence unit considered—Core, Supplement, and Excluded.

Its purpose is to:

- record the full search universe,

- preserve the rationale for tiering decisions,

- expose ambiguity and reporting gaps,

- provide a transparent boundary between usable and unusable evidence,

- prevent retroactive reinterpretation.

Each row of the Study Log includes the following required fields:

**Mandatory Fields**

| Field | Description |
|---|---|
| **Study_ID** | Unique identifier for the evidence unit. |
| **Design_Type** | Abstract design classification (comparative, predictive, structural). |
| **Tier_Assignment** | Core / Supplement / Excluded. |
| **Reasons_for_Tiering** | Justification referencing construct, measurement, and design alignment. |
| **Bias_Considerations** | Potential bias sources and their directional implications. |
| **Measurement/Definition Issues** | Any mismatch between construct and operationalization. |
| **Notes** | Clarifications, unresolved ambiguities, or flags for later review. |

**Function**

The Study Log serves as the project's memory system.

Without it, tiering decisions cannot be reconstructed, verified, or challenged.

In RECAP, undocumented decisions are treated as *methodological nullities*.

**8.2 Tier Table (Alignment Backbone)**

The Tier Table aggregates only evidence units that participate—directly or conditionally—in inference:

- Tier Core
- Tier Supplement

Excluded units *never* appear in the Tier Table.

**Mandatory Fields**

| Field | Description |
| --- | --- |
| **Methods Summary** | Concise abstraction of design, timing, and analytic structure. |
| **Evidence Type** | Role in inference: comparative estimate, predictive component, structural constraint, etc. |
| **Strengths** | Features that justify Core placement or contextual utility in Supplement. |
| **Limitations** | Methodological constraints (measurement mismatch, reporting gaps, design restrictions). |

**Function**

The Tier Table provides the **interpretive backbone** that links evidence to:

- the inferential route,
- the structural assumptions,
- and the epistemic boundaries of the project.

It clarifies **why** each study is used and **how** it contributes.

A project without a Tier Table—and without explicit evidence-role assignments—cannot demonstrate structural coherence.

**8.3 Reviewer Block + Analytic Memo**

*(Layer B: Structured Internal Scrutiny)*

The Reviewer Block simulates the reasoning of a rigorous external reviewer and formalizes Layer B within the RECAP architecture.

It has two complementary components:

1. **Reviewer Block** — external scrutiny emulation
2. **Analytic Memo** — internal reasoning trace

These components ensure that the inferential claims match the evidence and that assumptions are explicitly justified rather than absorbed implicitly into analysis.

**Reviewer Block — Required Elements**

Each project must articulate the following five items:

1. **Two methodological findings**

- Structural observations about the evidence base (not results).
- E.g., construct heterogeneity, measurement compression, design incompatibility.

2. **One conceptual insight**
    - A refinement or clarification related to constructs or the domain abstraction layer.
    - This may be eligible for upward transmission.

3. **One anticipated reviewer critique**
    - A plausible methodological or conceptual challenge and its pre-emptive response.
    - Must reference specific tiering or routing decisions.

4. **One disconfirming model**
    - A coherent alternative interpretation that could weaken or invert the claim.
    - This requirement implements falsificationist reasoning at the project level.

5. **Route assumptions (explicit list)**
    - All assumptions required for the chosen inferential route.
    - Each assumption must be paired with:
        - plausibility assessment,
        - potential failure modes,

- consequences for primary inference.

**Analytic Memo — Required Narrative**

The memo is a free-form but disciplined explanation of:

- how evidence was interpreted under the route's assumptions,
- where uncertainty resides,
- how boundary conditions were evaluated,
- how Supplement units informed robustness,
- how the project maintained compliance with RECAP's inheritance rules.

**Function**

The Analytic Memo is the **epistemic audit trail**.

It documents how reasoning unfolded and why the conclusions are proportional to the evidence.

Projects lacking an Analytic Memo have no reconstructable logic—and therefore cannot be said to be operating under RECAP.

**8.4 System-Level Rationale for Mandatory Outputs**

The required outputs serve three architectural purposes:

**1. Governance**

They enforce methodological discipline and prevent drift within and across projects.

**2. Transparency**

They make every inferential decision visible, defensible, and reproducible.

**3. Inheritance Compatibility**

They provide the structured information necessary for:

- upward transmission of abstract insight,

- backward tracing of contamination events,

- longitudinal coherence across programmatic research.

Mandatory outputs are thus not optional components—they are the *institutional infrastructure* of RECAP-as-a-system.

**9. Multi-Layer Inheritance Architecture**

The RECAP Framework functions as a *governed evidence ecosystem* rather than a project-level workflow.
Its coherence depends on a formal inheritance architecture that regulates how constraints and insights move across layers, and how projects remain insulated from one another while still benefiting from stable methodological foundations.

This architecture is not an analogy—it is a **mechanism**.
Without inheritance rules, RECAP collapses into the same single-layer logic that produces drift, contamination, and inconsistent reasoning across research programs.

RECAP defines **three non-interchangeable layers**:

1. **Grandparent — Meta-Engine**

2. **Parent — Domain Abstraction Layer**

3. **Child — Project Modules**

Each layer performs a distinct epistemic function, operates on different kinds of information, and evolves on different time scales.

The separation of these layers is the foundation that allows RECAP to scale across domains, investigators, and generations of research.

**9.1 Grandparent Layer — The Meta-Engine**

The **Grandparent** is RECAP's constitutional layer.

It encodes the methodological laws that govern *all* downstream reasoning.

**Characteristics**

1. **Domain-agnostic**

   No disease names, phenotypes, exposures, measurement tools, or design conventions may enter this layer.

2. **Permanently abstract**

   Its content consists solely of methodological laws—not examples, not heuristics, not exceptions.

3. **Immutable from below**

   Neither the Parent nor any Child module may modify, override, reinterpret, or "soften" Grandparent laws.

4. **Slow-changing but not frozen**

   It may evolve **only** through upward transmission of methodological insight that has been stripped of all context.

**Function**

The Grandparent layer defines:

- what counts as a construct,
- how inference must be structured,
- the one-route principle,
- the legality of upward vs downward information flow,
- the architecture of tiering, routing, and contamination control.

Without a Grandparent layer, RECAP would be unable to support cumulative methodological development; every new project would effectively begin from zero.

The Grandparent layer is the **meta-theoretical spine** of the entire system.

**9.2 Parent Layer — Domain Abstraction Layer**

The **Parent** layer translates Grandparent laws into structures appropriate for a specific domain (e.g., psychiatric phenotypes, ophthalmologic risk models, metabolic constructs). It is *abstract*, but not empty.

**What the Parent Layer May Contain**

- Domain-level constructs (e.g., "affective instability," "optic nerve vulnerability")
- Abstract measurement classes ("physiological indicators," "behavioral proxies")
- Permissible design forms (e.g., "longitudinal comparative structures," "quasi-experimental contrasts")

- The allowable inferential space for the domain

Importantly, these elements:

- are **not** tied to any specific dataset,
- are **not** tied to any specific operationalization,
- are **not** allowed to incorporate project-level assumptions or measurement conventions.

**What the Parent Layer May *Not* Contain**

- Study-specific variables
- Domain content expressed through concrete operational definitions
- Project-level terminology
- Measurement rules from Child modules
- Statistical model preferences
- Effect size conventions

**Function**

The Parent layer:

- provides a reusable structural scaffold for all Child modules in that domain,
- prevents projects from redefining constructs to suit convenience,
- enables consistency across projects and across time,
- ensures all children inherit the same interpretive boundaries.

It is the layer that makes a "research program" possible.

**9.3 Child Layer — Project Modules**

The **Child** layer is where empirical reasoning occurs.

Every Child module is a *complete but constrained* analytic universe that inherits rules from above and implements them without modification.

Each Child must independently specify:

- its evidence units,
- its single inferential route,
- its tiering assignments,
- its Reviewer Block and Analytic Memo,
- its contamination checks,
- its versioned decisions.

**Key Characteristics**

1. **Full isolation**
   Child modules cannot borrow assumptions, heuristics, measurement rules, or analytic preferences from one another.
   This is essential to prevent cross-project drift.

2. **One-route constraint**
   Each Child embodies exactly one inferential logic.

3. **Local flexibility**

Children may choose models, estimators, priors, statistical workflows, etc., *as long as* these choices do not alter Parent or Grandparent structure.

4. **Temporary lifespan**

   Children are not permanent institutions—they exist only for their project and leave behind audit trails, not structural laws.

**Function**

The Child layer is the **action layer** of RECAP:

it produces empirical claims while remaining fully constrained by inherited rules.

Its outputs can feed insight upward—but only when abstracted.

**9.4 Directional Inheritance Rules**

Inheritance in RECAP is *directional and asymmetric*.

**Permitted Flows**

**Grandparent → Parent**

Universal methodological laws flow downward.

**Parent → Child**

Domain abstractions flow downward to govern project behavior.

**Child → Parent (insight only)**

Only *abstractable methodological insight* may flow upward.

This flow is tightly regulated:

- it must be context-independent,

- it must be re-expressed in domain-free terms,
- it must undergo contamination screening.

**Prohibited Flows**

❌ **Parent → Grandparent**

Domain content cannot alter methodological law.

❌ **Child → Parent / Grandparent (content)**

Measurements, datasets, analytic heuristics, convenience assumptions cannot flow upward.

❌ **Child ↔ Child**

No lateral transfer of assumptions or conventions.

This architecture enforces inheritance discipline—one of RECAP's most important contributions.

**9.5 Why a Multi-Layer Architecture Enables Stability**

Traditional evidence systems collapse abstraction, domain, and project decisions into a single layer.

As a result:

- contextual decisions masquerade as universal rules,
- project choices contaminate domain definitions,
- domain heuristics distort methodological principles,

- interpretive drift accumulates invisibly over time.

RECAP prevents these failures by distributing responsibilities across layers that evolve on different temporal and conceptual scales:

- **Grandparent:** evolves slowly (methodological insight)
- **Parent:** evolves moderately (domain abstraction)
- **Child:** evolves rapidly (project implementations)

By separating *what must remain stable* from *what may vary*, RECAP achieves:

- structural coherence across generations of research,
- insulation against contextual creep,
- a lawful mechanism for refinement,
- compatibility with automation at scale,
- interpretability that persists across investigators.

This architecture transforms evidence synthesis from a series of disconnected projects into a **governed epistemic system**.

## 10. Non-Contamination Rules (Legal Code Version)

The RECAP Framework is defined not only by its tiering and routing systems, but by a **constitutional layer of non-contamination rules** that regulate the movement of information across the Grandparent, Parent, and Child layers. These rules are necessary for RECAP to function as a governed, inheritance-based evidence architecture. Any violation dissolves the structural guarantees that make multi-project coherence possible.

The rules below are expressed in legal-code form to remove interpretive ambiguity.

**Rule 1 — No Upward Content Flow**

**Domain content may not enter the Grandparent layer under any circumstances.**

**1.1 Prohibited upward elements**

The following classes of information are categorically barred from upward transmission:

- domain terminology,

- measurement structures,

- dataset-specific assumptions,

- project-specific inferential heuristics,

- empirical conventions derived from any Child module.

**1.2 Rationale**

Upward content flow permanently distorts the meta-engine, collapsing abstraction into specificity. Once this occurs, the Grandparent layer can no longer govern multiple domains, eliminating the possibility of a universal methodological framework.

**Rule 2 — No Downward Rewriting of Laws**

**Neither the Parent layer nor any Child module may modify, reinterpret, soften, override, or create exceptions to Grandparent-level methodological laws.**

**2.1 Examples of violations**

- Relaxing the one-route constraint for convenience.

- Allowing mixed construct definitions within a project.

- Treating measurement tools as construct-defining entities.

- Replacing disconfirming-model requirements with narrative justification.

**2.2 Rationale**

Downward rewriting fragments the inferential system, producing locally coherent but globally incompatible projects. Without this rule, no cumulative methodological refinement is possible.

**Rule 3 — No Horizontal Assumption Borrowing**

**Assumptions, heuristics, or measurement conventions may not migrate from one Child module to another, even when the domains appear similar.**

**3.1 Prohibited transfers**

- Borrowing construct definitions from a sibling project.

- Importing measurement choices because "they worked previously."

- Reusing bias adjustments or modeling conventions without re-justification.

- Allowing terminology from one project to set defaults in another.

**3.2 Rationale**

Horizontal borrowing is the most common source of long-term drift in multi-project programs. It produces untraceable dependencies and breaks the independence required for methodological inheritance.

**Rule 4 — Explicit Boundary Contracts**

**Every interaction between layers must occur through an explicit, auditable boundary contract.**

**4.1 Required elements of a boundary contract**

A valid contract must specify:

1. the type of information crossing the boundary,
2. the layer of origin and destination,
3. the legal justification for transfer,
4. the prohibition on implicit reinterpretation,
5. the mechanism for documenting the transfer.

**4.2 Rationale**

Boundary contracts prevent the subtle, unrecorded seepage of assumptions across layers, which is the primary mechanism through which contamination becomes invisible.

**Rule 5 — Meta-Engine Must Remain Abstract and Insulated**

**Only abstract methodological insight may enter the Grandparent layer, and only after being stripped of all domain content.**

**5.1 Conditions for legitimate upward insight**

An upward transmission is permitted only when:

- the proposed insight is independent of any measurement, dataset, or domain;

- the insight can be stated entirely in the conceptual vocabulary of the receiving layer;

- the insight is demonstrably methodological rather than contextual.

### 5.2 Rationale

This rule maintains the universal character of the meta-engine. Without insulation, the Grandparent layer becomes a repository of historical contingencies rather than a governing methodological constitution.

### 10.1 Enforcement

The enforcement regime is structural, not discretionary.

A module that violates any rule above—regardless of how precisely it tiers evidence or documents its route—has **exited the RECAP Framework**. RECAP is defined by its inheritance architecture; violations negate its guarantees of stability, transparency, and drift resistance.

### 10.1.1 Mandatory documentation of violations

Every contamination event must be logged.

Documentation must include:

1. **Rule violated**

2. **Nature of the contamination** (content, assumption, measurement, or structural)

3. **Downstream inferential risks**

4. **Corrective action taken**

5. **Versioned update** (timestamp + rationale)

These records are entered into the **Study Log** (for project-level violations) or the **Parent Log** (for domain-level violations). They remain part of the permanent genealogy of the framework.

### 10.1.2 Enforcement philosophy

Enforcement is not punitive.

Its purpose is:

- to preserve the integrity of the meta-engine,
- to ensure transparency of reasoning across generations,
- to maintain reconstructability of all decisions,
- to protect against drift in multi-project evidence ecosystems.

Contamination is treated as a structural hazard, not a moral failing.
The enforcement protocol is a safeguard for the framework's long-term viability.

## 11. Automation Potential

Automation is compatible with RECAP not because it expands the framework but because RECAP constrains it. Automation operates only as an executor of rules; it may accelerate implementation, but it may not alter the structure that governs implementation. Within RECAP, automation is therefore a subordinate process whose legitimacy depends entirely on its adherence to inheritance laws, contamination boundaries, and construct–measurement separation.

Automation is neither an epistemic agent nor a methodological authority. It cannot interpret constructs, cannot adjudicate tiering ambiguity, and cannot generate inferential claims. RECAP treats automation as a procedural extension of the Child layer—permitted to execute but prohibited from defining or modifying. This distinction is foundational: RECAP governs automation, not the reverse.

**11.1 Why Automation Fits the RECAP Architecture**

Automation is feasible within RECAP because the framework explicitly defines:

1. construct boundaries,
2. permissible measurement correspondences,
3. tiering criteria and routing logic,
4. directional rules for information flow.

These specifications create a stable interface that automated systems can implement without introducing conceptual drift. Under RECAP, automation may:

- assist in tier assignment by applying human-defined criteria,
- enforce the one-route constraint by detecting hybrid reasoning,
- identify contamination events by monitoring prohibited flows,
- generate Study Log and Tier Table entries with complete provenance,
- support version-controlled updates across Child modules.

Automation succeeds because RECAP specifies the limits within which automation must remain. The meta-engine provides structure; automation provides speed.

## 11.2 Limits of Automation (Non-Negotiable Constraints)

Automation faces categorical restrictions under RECAP. These are structural, not pragmatic.

Automation **cannot**:

1. construct, revise, or reinterpret Grandparent-level laws;

2. generate new domain abstractions without explicit Parent-layer authorization;

3. resolve construct–measurement misalignment requiring conceptual judgment;

4. override tiering or routing decisions;

5. introduce new assumptions, heuristics, or definitions into any layer;

6. perform upward insight transmission;

7. participate in contamination, whether upward, downward, or horizontal.

Automation is confined to procedural execution at the Child level.
All epistemic authority remains human.

The Grandparent layer is permanently insulated from automated modification.
The Parent layer may incorporate automated pattern recognition only after human validation.
The Child layer permits automation most freely but remains rule-bound.

## 11.3 Conditions for Valid Automation within RECAP

Automation is admissible only if it satisfies all three structural conditions.

**Condition 1 — Explicit Rule Encoding**

Every automated action must map to a documented RECAP rule.

Opaque heuristics, stochastic processes, or model-derived shortcuts that cannot be traced to specific constraints are prohibited.

Automation may execute rules but may not author them.

**Condition 2 — Versioning and Auditability**

All automated decisions must be fully auditable, including:

- inputs,

- intermediate states,

- outputs,

- timestamps,

- software version and parameters used.

Silent modification of tier assignments, route specifications, or construct definitions is not permitted.

Automation must record its operations such that any decision can be reconstructed and challenged.

**Condition 3 — No Automated Inference**

Automation may summarize, categorize, or detect patterns, but it may not:

- interpret constructs,

- make inferential claims,

- determine conceptual boundaries,

- adjudicate theoretical alternatives,

- generate disconfirming models.

Inference is inherently conceptual and belongs to human reasoning within Layers A and B.

Automation may not engage in interpretation; it may only process.

### 11.4 Automation as a Child-Layer Executor

Under RECAP, automation is formally classified as a Child-layer function.

It:

- receives constraints from Parent and Grandparent layers,

- cannot transmit content upward,

- may only execute downward-propagated rules,

- must preserve tiering, routing, and contamination laws.

Automation thus becomes structurally safe: it accelerates execution without altering method.

This relationship is a defining feature of RECAP's architecture and distinguishes the framework from conventional AI-assisted workflows.

### 11.5 Structural Justification for Automation Limits

The limits above are not technical—they are philosophical.

They prevent the following risks:

1. **Reification** — allowing automated features to redefine constructs.

2. **Assumption Drift** — allowing models to propagate domain interpretations across projects.

3. **Rule Overwriting** — permitting algorithms to modify or suppress explicit methodological laws.

4. **Epistemic Delegation** — outsourcing conceptual reasoning to systems that lack interpretive capacity.

5. **Silent Contamination** — introduction of unauthorized assumptions across layers or projects.

By bounding automation, RECAP ensures that evidence systems remain interpretable across generations and that methodological stewardship remains human.

## 11.6 Summary

Automation, within RECAP, is powerful but fundamentally limited.
It can increase speed, reduce manual burden, and support large-scale evidence ecosystems, but it cannot:

- interpret,
- decide,
- infer,
- redefine,
- justify,

- or govern.

Automation is procedural; RECAP is structural.

The architecture ensures that automation enhances execution without compromising conceptual clarity, methodological rigor, or the integrity of the inheritance system.

## 12. Toy Example: A Multi-Layer Instantiation Using Abstract Constructs

This section illustrates how RECAP operates using abstract constructs (A, B, C) without domain meaning.

The example demonstrates tiering, routing, inheritance, contamination governance, and required study-level outputs.

### 12.1 Scenario

Suppose three studies examine relationships among three constructs:

- **A** (an abstract exposure construct)
- **B** (an intermediate construct)
- **C** (an outcome construct)

Each study operationalizes these constructs through different measurement procedures (m1, m2, m3).

The goal is not to estimate effects but to illustrate how RECAP organizes evidence.

### 12.2 Tiering

**Study S1**

- Measurement of A via m1 aligns well with its construct definition.
- B is measured using a partial proxy with known limitations.
- C is clearly defined and transparent.
    - → **Tier Core** (limitations noted)

**Study S2**

- A and B measured ambiguously.
- Construct definitions partially missing.
    - → **Tier Supplement** (cannot anchor inferential claims but useful for sensitivity)

**Study S3**

- Measurements do not correspond to constructs A or B.
- Key design elements unreported.
    - → **Tier Excluded** (recorded but not used)

Tiering output (Study Log excerpt):

| Study_ID | Tier | Reason_for_routing | Measurement_Issues |
|---|---|---|---|
| S1 | Core | Construct alignment | Proxy for B |
| S2 | Supplement | Partial mismatch | Ambiguous A/B |
| S3 | Excluded | Definition opacity | Non-correspondence |

**12.3 Routing**

Each Core or Supplement study must be assigned **exactly one route**, based on inferential role.

Possible abstract routes:

- **Comparative estimation (R1)**
- **Associational estimation (R2)**
- **Measurement evaluation (R3)**
- **Predictive modeling (R4)**

Routing for example studies:

- S1 → **R2** (associational)
- S2 → **R3** (measurement evaluation)

Routing Table excerpt:

| **Study_ID** | Tier | Assigned Route | Evidence Type |
|---|---|---|---|
| S1 | Core | R2 | Asscotion |
| S2 | Supplement | R3 | Measurement |

No study may take multiple routes.

This prevents epistemic overreach.

### 12.4 Parent and Child Instantiation

**Grandparent Layer**

Declares

- Construct definitions for A, B, C
- Tiering rules
- Contamination laws
- One-route constraint

This layer remains permanent and abstract.

**Parent Layer (Domain Abstraction)**

Translates A/B/C into domain-level abstractions (still abstract here).

Specifies measurement correspondence rules (m1 ↔ A, m2 ↔ B, etc).

**Child Project**

Implements a question such as:

**"How does A relate to C, mediated through B?"**

Uses Parent-level rules to classify S1 and S2.

Generates outputs (Study Log, Tier Table, Reviewer Block).

**12.5 Contamination Governance in the Example**

**Upward Contamination Check**

- S1 uses m1 to measure A.
- m1 is *not* allowed to redefine the construct A in Grandparent.
- → Blocked.

**Downward Contamination Check**

- Child project cannot rewrite the definition of A or B.
- No violation.

**Horizontal Contamination Check**

- S1's proxy assumption for B cannot be reused by S2.
- → Prevented.

Boundary Contract Summary:

| Layer | Allowed | Prohibited |
|---|---|---|
| Grandparent | Laws, insight | Domain content |
| Parent | Abstractions | Law rewriting |
| Child | Application | Cross-study assumption borrowing |

## 12.6 Required Outputs

**Study Log Row (S1 example)**

- ID: S1
- Design Type: Observational (Abstract)
- Tier: Core
- Bias Considerations: Proxy for B → nondirectional risk
- Measurement Issues: Partial misalignment for B
- Notes: Adequate for R2

**Tier Table Row (S1)**

- Methods Summary: Association between A and C via m1–m3 mapping
- Evidence Type: Associational
- Strengths: Clear construct A; transparent m1
- Limitations: Proxy for B

**Reviewer Block (S1)**

- Two methodological findings:
    1. Construct A measured reliably.
    2. Proxy B introduces potential attenuation.
- Conceptual insight: Operationalization of B remains unstable.
- Reviewer critique: Why was a stronger proxy not used?
- Disconfirming model: C may influence B rather than vice versa.
- Assumptions: m1 valid for A; proxy B monotonic.

## 12.7 Interpretation

This toy example demonstrates that RECAP:

1. Separates construct layers from measurement.
2. Enforces tiering and routing rigor.
3. Prevents contamination across studies and layers.
4. Produces auditable, reproducible study-level outputs.
5. Supports an expandable evidence ecosystem without drift.

This example is intentionally domain-free to show that RECAP operates independently of disciplinary context.

## 13. Versioning and Governance

RECAP v1.0 constitutes the inaugural specification of a methodological paradigm that did not previously exist in evidence science.
It defines the foundational laws, inheritance constraints, transparency requirements, and contamination rules upon which all subsequent versions must rely.
The Grandparent layer anchors these principles permanently; later versions may extend but may not rewrite these laws.

Versioning in RECAP follows a governance architecture analogous to specification-based systems in other formal disciplines.
A new version is justified only when a methodological insight—rather than a domain convention or empirical result—demonstrates a structural limitation in the current model.
Insights must be abstract, upward-transmissible, and independent of any specific Parent or Child instantiation.

### 13.1 What Qualifies as a Legitimate Upgrade

A version increment (v1.x → v1.y or v2.0) requires evidence that:

1. **A Grandparent-level law is incomplete or insufficiently general**, demonstrated through repeated methodological challenges across multiple projects.
2. **A new principle improves drift-resistance or enhances cross-project coherence**, without introducing domain content into the meta-engine.
3. **The inheritance architecture can be strengthened without violating existing boundary contracts.**

Version upgrades may refine:

- contamination governance,
- routing logic,
- tiering criteria,
- explicit boundary contracts.

They may not modify:

- anti-reification principles,
- one-route constraint,
- separation of constructs and measurements,
- downward insulation of the Grandparent layer.

These restrictions preserve RECAP as a long-term methodological lineage rather than a mutable toolkit.

## 13.2 Inheritance of Versions Across Layers

The Grandparent layer determines the version of the entire system.

Parent layers inherit the current version as constraints, and may append domain abstraction rules without altering inherited laws.

Child projects inherit from both and remain bound to all non-contamination rules.

A Parent engine may declare its own version (P1.0, P1.1, etc.) but such versions represent domain-level elaborations, not alterations of Grandparent laws.

Child-level versioning is optional and used primarily for automation debugging or audit trails.

## 13.3 Governance Structure

RECAP governance rests on three principles:

### 1. Irreversibility of Laws

Grandparent-level laws can be expanded but not rescinded.

This protects the architecture from conceptual drift and ensures interpretability across generations.

### 2. Transparency of Change

Every version increment must include a formal changelog documenting:

- the insight motivating the change,
- the boundary it affects,

- the reasoning for its generalizability.

**3. Methodological Justification**

Changes cannot arise from efficiency, convenience, or automation needs;

only methodological reasoning is admissible.

This governance model ensures that RECAP evolves through philosophical rigor rather than ad hoc adaptations.

**13.4 RECAP v1.0 as the Origin Point of a Lineage**

To our knowledge, no prior framework has formalized a multi-layer inheritance structure for evidence systems, nor articulated the laws required to preserve separation between abstract method, domain abstraction, and project instantiation.
RECAP v1.0 therefore represents the origin point of a methodological lineage intended to support multi-decade development.

Future versions may extend its scope, elaborate Parent-engine templates, or integrate responsibly with automated systems,
but all such expansions inherit the foundational laws documented here.

**14. Comparison to Existing Frameworks**

Existing frameworks in evidence science provide essential contributions but operate within a single-layer architecture.

They guide reporting, assess bias, model causal relationships, or standardize certainty grading, yet none define a governance layer that regulates cross-project inheritance, contamination control, and the stability of methodological laws.

RECAP does not compete with these systems; it governs the abstraction level above them.

### 14.1 Reporting Guidelines (e.g., PRISMA, CONSORT)

Reporting frameworks establish transparency in documenting study procedures and synthesis workflows.

They do not define constructs, tiering rules, routing logic, or inheritance constraints.

They also do not protect against cross-project drift or methodological contamination.

RECAP can incorporate PRISMA as an implementation detail within Child modules while providing meta-governance above it.

### 14.2 Risk-of-Bias and Quality Assessment Tools (e.g., ROB2, QUIPS)

These tools evaluate threats to validity within individual studies.

They do not manage relationships across studies, nor do they define how methodological decisions propagate across projects or generations.

RECAP complements these tools by specifying where such assessments reside in the multi-layer architecture and how they influence tiering and routing.

## 14.3 Causal Inference Frameworks (e.g., Potential Outcomes, DAGs)

Causal frameworks govern identification, counterfactual reasoning, and causal structure representation.

They do not govern multi-project evidence ecosystems, nor do they address contamination between measurement rules, constructs, or domain abstractions across studies.

RECAP is agnostic to causal modeling: causal models may operate within Child modules, but RECAP governs the methodological context in which such models are instantiated.

## 14.4 Certainty and Grading Frameworks (e.g., GRADE)

Certainty frameworks grade the strength of bodies of evidence.

They do not impose multi-layer inheritance rules or restrict how constructs, measurements, or assumptions may flow across studies or generations.

RECAP provides a structural layer above GRADE, determining how graded evidence fits within tiering, routing, and boundary contracts.

## 14.5 Meta-analytic and Statistical Models

Meta-analytic methods integrate estimates across studies but do not define how studies are tiered, routed, or governed prior to aggregation.

They assume—not specify—the architecture in which integration occurs.

RECAP defines that architecture, ensuring that statistical models operate within a stable inferential environment.

### 14.6 AI-based Evidence Systems

Automated tools can accelerate extraction, study classification, and analysis.

However, AI systems lack a meta-engine to determine:

- which rules are universal,
- which decisions are project-specific,
- how insights propagate,
- how contamination is prevented.

RECAP offers the governance framework that AI systems require for stable, transparent, and interpretable deployment.

### 14.7 Summary of Distinctions

| Feature | Existing Frameworks | RECAP |
|---|---|---|
| **Governs reporting** | Yes | **Indirect (via Child)** |

| Governs bias tools | No | Yes (via tiering/routing) |
| --- | --- | --- |
| Governs causal models | No | Yes (as instantiations) |
| Multi-project coherence | No | Yes |
| Inheritance architecture | No | Yes |
| Contamination rules | No | Yes |
| Transparency outputs | Partial | Mandatory |
| Drift resistance | No | Yes |
| Layer separation | No | Yes |

## 14.8 Paradigm Placement

To our knowledge, no existing system operates at the abstraction level introduced by RECAP.

Where prior frameworks address components of evidence synthesis, RECAP defines the **governing architecture** within which these components must reside.

This distinction is foundational: RECAP establishes a new conceptual layer above reporting, bias assessment, modeling, and grading frameworks.

## 15. Limitations & Misuse Risks

Although RECAP introduces a new abstraction layer for evidence governance, its scope and constraints must be clearly delineated to prevent misuse.

This section outlines limitations inherent to the framework and risks arising from incorrect implementation.

## 15.1 Conceptual Overreach

RECAP governs methodological structure, not substantive inference.

Users may mistakenly treat tiering, routing, or contamination laws as substitutes for domain expertise.

RECAP does not eliminate the need for critical interpretation, theoretical reasoning, or context-specific judgment.

## 15.2 Risk of Overformalization

The architecture imposes structure that may feel excessive for small, one-off projects.

In such settings, multi-layer governance may add complexity without proportional gain.

RECAP is most effective when multiple projects, teams, or research generations must align under a unified framework.

## 15.3 Dependence on Construct Clarity

RECAP requires constructs to be explicitly defined prior to measurement alignment.

If researchers provide vague or unstable constructs, the framework can only identify deficiencies, not repair them.

Failure to articulate constructs limits the validity of tiering and routing decisions.

## 15.4 Misinterpretation as a Checklist

RECAP is an architecture, not a procedural checklist.

Improper use may reduce its principles to mechanical scoring systems or rigid templates.

Such reduction undermines the purpose of the framework: preserving inferential integrity across layers.

## 15.5 Insufficient Governance at the Grandparent Level

The permanent meta-engine requires careful stewardship.

If future versions are modified without methodological justification—or by domain-driven pressure—the framework may drift away from its philosophical foundations.

This risk is mitigated through version control and strict upward-transmission rules but cannot be eliminated entirely.

## 15.6 Automation Misuse

Automation may enforce RECAP but cannot interpret constructs or provide conceptual insight.

If automated systems attempt to rewrite Grandparent laws, inferential reasoning may collapse into opaque, model-driven heuristics.

AI tools must be explicitly constrained by boundary contracts and audit requirements.

### 15.7 Boundary Violations and Contamination Risks

Users may inadvertently:

- import assumptions across Child modules (horizontal contamination),
- rewrite Parent abstractions using project-level conventions (downward contamination), or
- elevate domain content into the Grandparent layer (upward contamination).

Each violation erodes the separation of layers and compromises the long-term stability of the evidence ecosystem.

### 15.8 Incomplete Adoption Across Teams

The architecture assumes that all contributors accept the same boundary contracts.
If teams implement RECAP inconsistently—especially in multi-institution collaborations—the coherence and drift-resistance of the system may weaken.
Partial adoption is possible, but full benefit requires systemic implementation.

## 15.9 Paradigm Misplacement

RECAP governs evidence organization, not statistical modeling, causal identification, decision theory, or clinical guidelines.

Incorrectly applying RECAP to these areas may blur methodological distinctions and dilute the framework's effectiveness.

## 15.10 Summary

RECAP is powerful but demands disciplined implementation.

Its value emerges in proportion to construct clarity, adherence to contamination laws, stewardship of the Grandparent layer, and appropriate integration with domain and project-level reasoning.

Misuse risks arise not from the architecture itself but from deviations in practice or attempts to repurpose RECAP for roles it was not designed to fulfill.

## 16. Future Research Program

RECAP v1.0 establishes the origin point of a methodological lineage.

This section outlines the long-term research program that may extend the framework across domains, generations, and automated systems while preserving its governing laws.

### 16.1 Development of Domain-Specific Parent Engines

Parent engines translate Grandparent laws into domain-level abstractions.

Future work includes designing structured templates for these engines, specifying:

- construct definitions appropriate for each domain,
- measurement correspondence rules,
- domain-specific boundary contracts.

Each Parent engine (P1.0, P2.0, …) becomes a bridge between universal laws and concrete implementation.

### 16.2 Automated Child Generators

Because RECAP separates rules from content, automated systems may one day generate Child modules directly from study corpora.

Work is needed to:

- encode tiering and routing rules,
- detect contamination risks,
- generate Study Log and Tier Table outputs automatically,
- integrate adversarial reasoning into Reviewer Blocks.

Such systems would accelerate evidence synthesis while preserving the integrity of the inheritance architecture.

### 16.3 Cross-Domain Orchestration Systems

Multiple Parent engines may coexist and interact across complex interdisciplinary settings.

A future research direction is the development of orchestration layers that:

- coordinate constructs across domains,
- maintain boundary separation,
- detect cross-domain contamination,
- support shared philosophical constraints.

This enables structured reasoning across heterogeneous evidence spaces.

### 16.4 Long-Term Governance and Stewardship

The meta-engine will require sustained philosophical and methodological governance.

Future research may refine:

- changelog standards,
- version control mechanisms,
- processes for admitting upward insight,
- criteria for Grandparent-level revision.

Stewardship ensures stability across decades of use.

### 16.5 Drift Analysis and Resilience Testing

RECAP asserts that contamination control prevents drift across research generations.

Empirical and simulation-based studies may test:

- the degree to which drift is reduced,
- scenarios in which contamination pressures accumulate,
- resilience of tiering and routing under inconsistent adoption.

Such work will evaluate the durability of the architecture under real-world conditions.

---

## 16.6 Expansion of Transparency Outputs

Study Log, Tier Table, and Reviewer Block represent the initial output set.

Future work may expand:

- machine-readable metadata,
- audit trails for automated systems,
- cross-version interpretability layers,
- visualization tools for multi-layer inheritance.

These outputs aim to embed transparency into the architecture itself.

## 16.7 Embedding RECAP into Educational Frameworks

Because RECAP introduces a new abstraction layer for evidence reasoning, education systems may need to incorporate:

- tiering logic,
- boundary contracts,
- inheritance architecture,
- contamination governance.

This transforms how evidence is taught and interpreted.

**16.8 Establishing the RECAP Family of Frameworks**

The long-term trajectory of RECAP is the formation of a family of derivative frameworks, each inheriting v1.0 laws while contributing domain-level insights. Examples include:

- Parent engines for specific fields,
- Child generators for different evidence modalities,
- Orchestration systems across domains.

These expansions preserve alignment while enabling innovation.

**16.9 Summary**

RECAP v1.0 defines a starting point—not an endpoint.

Its purpose is to anchor a multi-decade evolution of evidence architecture, governed by explicit laws, inheritance rules, and boundary contracts.

Future research will extend RECAP across domains, automate its execution, test its resilience, and cultivate the family of frameworks that will emerge from its foundations.

## 17. Conclusion

RECAP v1.0 introduces a formal abstraction layer above existing evidence frameworks, separating methodological law from domain content and project implementation.

By articulating a multi-layer inheritance architecture—Grandparent, Parent, and Child—it provides a foundation for long-term coherence, drift resistance, and structured evolution across research generations.

The framework's tiering rules, routing constraints, and contamination governance mechanisms create a disciplined environment in which inference can be organized, compared, and preserved with conceptual clarity.

These principles do not replace causal modeling, statistical methods, or reporting standards; rather, they govern the methodological space in which such tools operate.

RECAP's purpose is not to prescribe substantive conclusions, but to formalize the structural conditions under which evidence can accumulate reliably across heterogeneous projects.

By enforcing separation of layers, defining explicit boundary contracts, and mandating transparency outputs, RECAP establishes a stable meta-engine capable of supporting systematic, multi-decade development of evidence systems.

As the inaugural specification of this architecture, RECAP v1.0 marks the beginning of a structured research program.

Future versions may refine its laws, elaborate domain-specific Parent engines, and integrate automated Child generators, but all expansions inherit from the principles introduced here.

In this sense, RECAP serves as both a conceptual foundation and a governance mechanism for the evidence ecosystems that will emerge from it.

**C. Evidence-Synthesis Frameworks & Reporting Standards**

*systematic reviews of interventions* (version 6.3). Cochrane.

https://training.cochrane.org/handbook

### D. Bias Assessment Tools

**13. Sterne, J. A. C., Savović, J., Page, M. J., et al.** (2019). RoB 2: A revised tool for assessing risk of bias in randomised trials. *BMJ, 366*, l4898. https://doi.org/10.1136/bmj.l4898

**14. Hayden, J. A., van der Windt, D. A., Cartwright, J. L., Côté, P., & Bombardier, C.** (2013). Assessing bias in studies of prognostic factors. *Annals of Internal Medicine, 158*(4), 280–286. https://doi.org/10.7326/0003-4819-158-4-201302190-00009

### E. Meta-Research, Reproducibility, Drift

**15. Ioannidis, J. P. A.** (2005). Why most published research findings are false. *PLS Medicine, 2*(8), e124. https://doi.org/10.1371/journal.pmed.0020124

**16. Open Science Collaboration.** (2015). Estimating the reproducibility of psychological science. *Science, 349*(6251), aac4716. https://doi.org/10.1126/science.aac4716

### F. AI in Evidence Synthesis / Automation

**17. van de Schoot, R., de Bruin, J., Schram, R., et al.** (2021). An open source machine learning framework for efficient and transparent systematic reviews. *Nature Machine*

FIGURE LEGENDS

**Figure 1. RECAP Three-Layer Architecture**

This figure illustrates the hierarchical structure of the RECAP Framework, composed of three non-interchangeable layers.

The **Grandparent layer** defines universal methodological laws, including construct definitions, tiering rules, contamination constraints, and the one-route principle. Through **inheritance**, these constraints flow downward to the **Parent layer**, which establishes domain-level abstractions, measurement correspondence rules, and cross-project routing structure.

Through **application**, the **Child layer** instantiates specific projects, defining project-level questions, observed data, and required outputs (Study Logs, Tier Tables, Reviewer Blocks).

The arrows indicate **permitted information flow**; layers remain insulated from upward content flow, downward rewriting, and horizontal borrowing.

**Figure 2. Evidence Governance and Routing Workflow Under RECAP**

This flow diagram depicts the procedural governance structure that organizes evidence within a RECAP Child project.

The process begins with **construct definition** (Grandparent constraints) and **declaration of a single inferential route**, including assumptions, boundary conditions, and a disconfirming model.

Evidence units are then retrieved and processed through the **tiering algorithm**, which assigns each unit to Core, Supplement, or Excluded tiers based on construct alignment, measurement adequacy, and design transparency

Figure 2 — RECAP Evidence Gover.

Core units undergo a **routing-coherence check** to ensure compatibility with the designated route.

The route and tier assignments are then **frozen and versioned** in the Study Log.

A **contamination-governance overlay** enforces RECAP's inheritance rules, preventing upward, downward, or horizontal contamination across layers or projects.